\documentclass[12pt]{article}
\usepackage{epsf}
\title{ {\bf Lepton flavor violating $Z\rightarrow l_1^+ l_2^-$ decay in the
split fermion scenario in the two Higgs Doublet model}}
\author{\vspace{1cm}\\
        {\bf E. O. Iltan}
        \thanks{E-mail address:
        eiltan@heraklit.physics.metu.edu.tr}
 \\
        Physics Department, Middle East Technical University \\
        Ankara, Turkey\\}

\date{}

\begin{document}
\setlength{\baselineskip}{24pt}
\maketitle
\setlength{\baselineskip}{7mm}
\begin{abstract}
We predict the branching ratios of $Z\rightarrow e^{\pm}
\mu^{\pm}$, $Z\rightarrow e^{\pm} \tau^{\pm}$ and $Z\rightarrow
\mu^{\pm} \tau^{\pm}$ decays in the framework of the 2HDM, in the
split fermion scenario. We observe that the branching ratios are
not sensitive to a single extra dimension, however, this
sensitivity is considerably large for two extra dimensions.
\end{abstract}
\thispagestyle{empty}
\newpage
\setcounter{page}{1}
\section{Introduction}
The lepton flavor violating (LFV) Z decays are clean from the
theoretical point of view since they are free from the long
distance effects. On the other hand, they are rich in the sense
that they exist at least at one loop level and carry a
considerable information about the free parameters of the model
used. Therefore, it is worthwhile to analyze these decays and
there is an extensive work related to them in the literature
\cite{Riemann}-\cite{Perez}.

Since the lepton flavor is conserved in the SM, for the flavor
violation in the lepton sector, one needs to extend the SM. The so
called $\nu$SM model, which is constructed by taking neutrinos
massive and permitting the lepton mixing mechanism
\cite{Pontecorvo}, is one of the candidate. However, the
theoretical predictions for the branching ratios (BRs) of the LFV
Z decays in this model are extremely small in the case of internal
light neutrinos \cite{Riemann,Illana}
\begin{eqnarray}
BR(Z\rightarrow e^{\pm} \mu^{\pm})\sim BR(Z\rightarrow e^{\pm}
\tau^{\pm})
&\sim& 10^{-54}  \nonumber \, , \\
BR(Z\rightarrow \mu^{\pm} \tau^{\pm}) &<& 4\times 10^{-60} .
\label{Theo1}
\end{eqnarray}
and they are far from the experimental limits obtained at LEP 1
\cite{PartData}:
\begin{eqnarray}
BR(Z\rightarrow e^{\pm} \mu^{\pm}) &<& 1.7\times 10^{-6} \,\,\,
\cite{Opal}
\nonumber \, , \\
BR(Z\rightarrow e^{\pm} \tau^{\pm}) &<& 9.8\times 10^{-6}\,\,\,
\cite{Opal,L3} \nonumber \, , \\
BR(Z\rightarrow \mu^{\pm} \tau^{\pm}) &<& 1.2\times 10^{-5} \,\,\,
\cite{Opal,Delphi} \label{Expr1}
\end{eqnarray}
and from the improved ones at Giga-Z \cite{Wilson}:
\begin{eqnarray}
BR(Z\rightarrow e^{\pm} \mu^{\pm}) &<& 2\times 10^{-9}  \nonumber \, , \\
BR(Z\rightarrow e^{\pm} \tau^{\pm}) &<& f\times 6.5\times 10^{-8}
\nonumber \, , \\
BR(Z\rightarrow \mu^{\pm} \tau^{\pm}) &<& f\times 2.2\times
10^{-8} \label{Expr2}
\end{eqnarray}
with $f=0.2-1.0$. Notice that these numbers are obtained for the
decays $Z\rightarrow \bar{l}_1 l_2+ \bar{l}_2 l_1$, namely
\begin{eqnarray}
BR(Z\rightarrow l_1^{\pm} l_2^{\pm})=\frac{\Gamma (Z\rightarrow
\bar{l}_1 l_2+ \bar{l}_2 l_1)}{\Gamma_Z} \, .
\end{eqnarray}

Other possible scenarios to enhance the BRs of the corresponding
LFV Z decays are the extension of $\nu$SM with one heavy ordinary
Dirac neutrino \cite{Illana}, the extension of $\nu$SM with two
heavy right-handed singlet Majorana neutrinos \cite{Illana}, the
Zee model \cite{Ghosal}, the model III version of the two Higgs
doublet model (2HDM), which is the minimal extension of the SM
\cite{EiltZl1l2}, the supersymmetric models \cite{Masip, Cao},
top-color assisted technicolor model \cite{Yue}.

In this work, we study the LFV processes $Z\rightarrow e^{\pm}
\mu^{\pm}$, $Z\rightarrow e^{\pm} \tau^{\pm}$ and $Z\rightarrow
\mu^{\pm} \tau^{\pm}$ in the framework of the 2HDM in the split
fermion scenario \cite{Hamed}. This scenario is based on the idea
that the hierarchy of fermion masses are coming from the overlap
of the fermion Gaussian profiles in the extra dimensions and
reached great interest in the literature
\cite{Hamed}-\cite{Surujon}. In \cite{Mirabelli} the explicit
positions of left and right handed components of fermions in the
extra dimensions have been predicted. \cite{Changg} is devoted to
the restrictions on the split fermions in the extra dimensions
using the leptonic W decays and the lepton violating processes.
The CP violation in the quark sector has been analyzed in
\cite{Branco} and to find stringent bounds on the size of the
compactification scale 1/R, the physics of kaon, neutron and B/D
mesons  has been studied in \cite{Chang2}. In \cite{Hewett} the
rare processes in the split fermion scenario and in \cite{Perez2}
the shapes and overlaps of the fermion wave functions in the split
fermion model have been considered. In \cite{IltanEDMSplit} the
electric dipole moments of charged leptons have been predicted,
and in \cite{IltanRadLeptSplit} the radiative  LFV decays have
been examined, in the framework of this scenario. Recently, the
Higgs localization in the split fermion models has  been analyzed
in \cite{Surujon}.

In our calculations we estimate the sensitivities of BRs of the
LFV Z decays to the compactification scale and the Gaussian widths
of the charged leptons in the extra dimensions. We make the
analysis including a single extra dimension and observe that the
enhancement in the BRs of these decays are small. However, in the
case of two extra dimensions, especially the one that charged
leptons are restricted to the fifth extra dimension, with non-zero
Gaussian profiles, there is a considerable enhancement in the BRs
of the decays under consideration, even more than one order for
the compactification scale interval considered.

The paper is organized as follows: In Section 2, we present the
effective vertex and the BRs of LFV Z decays in the split fermion
scenario. Section 3 is devoted to discussion and our conclusions.
In appendix section, we give the explicit expressions of the form
factors appearing in the effective vertex.
\section{$Z\rightarrow l_1^- l_2^+$ decay in the split fermion
scenario in the two  Higgs doublet model.}
The extremely small theoretical values of the BRs of the LFV Z
boson decays forces one to go beyond and search a new mechanism to
enhance these numerical values near to the experimental limits.
From the theoretical point of view, the existence of the flavor
changing neutral currents (FCNCs) is essential to create the LFV
processes and, the multi Higgs doublet models, which are
constructed by extending the Higgs sector of the SM, are among the
possible models which permits the FCNC currents at tree level. The
2HDM is one of the candidate for the multi Higgs doublet model
and, in general, it permits the FCNC at tree level.  The LFV Z
decay $Z\rightarrow l_1^- l_2^+$ can be induced at least in the
one loop level in the framework of the 2HDM and new Higgs scalars
play the main role for the large BRs of these decays. Furthermore,
the inclusion of the spatial extra dimensions brings additional
contributions to the physical quantities of the decays under
consideration. Here, we respect the idea that the hierarchy of
lepton masses are coming from the lepton Gaussian profiles in the
extra dimensions, so called split fermion scenario.

The LFV Z decay $Z\rightarrow l_1^- l_2^+$ exist with the help of
the Yukawa interactions and, in a single extra dimension,
respecting the split fermion scenario, it reads
\begin{eqnarray}
{\cal{L}}_{Y}=
\xi^{E}_{5 \,ij} \bar{\hat{l}}_{i L} \phi_{2} \hat{E}_{j R} + h.c.
\,\,\, , \label{lagrangian}
\end{eqnarray}
where $L$ and $R$ denote chiral projections $L(R)=1/2(1\mp
\gamma_5)$, $\phi_{2}$ is the new scalar doublet and $\xi^{E}_{5\,
ij}$ are the flavor violating Yukawa couplings in five dimensions
and they are complex in general. The lepton fields $\hat{l}_{i L}$
($\hat{E}_{j R}$) are the zero mode \footnote{Notice that we take
only the zero mode lepton fields in our calculations.} lepton
doublets (singlets) with Gaussian profiles in the extra dimension
which is represented by the coordinate $y$ and they can be defined
as
\begin{eqnarray}
\hat{l}_{i L}&=& N\,e^{-(y-y_{i L})^2/2 \sigma^2}\,l_{i L} ,
\nonumber
\\ \hat{E}_{j R}&=&N\, e^{-(y-y_{j R})^2/2 \sigma^2}\, E_{j R}\, .
\label{gaussianprof}
\end{eqnarray}
where $l_{i L}$ ($E_{j R}$) are the lepton doublets (singlets) in
four dimensions with family indices $i$ and $j$, $\sigma$,
satisfying the property $\sigma << R$, is the parameter
representing the Gaussian width of the leptons in five dimensions
and N is the normalization factor, $N=\frac{1}{\pi^{1/4}\,
\sigma^{1/2}}$. Here the coordinates $y_{i L}$ and $y_{iR}$
represent the positions of the peaks of left and right handed
parts of $i^{th}$ lepton in the fifth dimension and they are
obtained by assuming that the mass hierarchy of leptons are coming
from  the relative positions of the Gaussian peaks of the wave
functions located in the extra dimension \cite{Hamed, Mirabelli}.
The observed lepton masses are the sources to calculate these
coordinates and in \cite{Mirabelli} one possible set of locations
of the lepton fields in the fifth dimension has been estimated as
\begin{eqnarray}
P_{l_i}=\sqrt{2}\,\sigma\, \left(\begin{array}{c c c}
11.075\\1.0\\0.0
\end{array}\right)\,,\,\,\,\, P_{e_i}=\sqrt{2}\,\sigma\, \left(\begin{array}
{c c c} 5.9475\\4.9475\\-3.1498
\end{array}\right)
 \,\, . \label{location}
\end{eqnarray}

Now, we would like the present the Higgs sector of the model under
consideration. The Higgs doublets $\phi_{1}$ and $\phi_{2}$ are
chosen as
\begin{eqnarray}
\phi_{1}=\frac{1}{\sqrt{2}}\left[\left(\begin{array}{c c}
0\\v+H^{0}\end{array}\right)\; + \left(\begin{array}{c c} \sqrt{2}
\chi^{+}\\ i \chi^{0}\end{array}\right) \right]\, ;
\phi_{2}=\frac{1}{\sqrt{2}}\left(\begin{array}{c c} \sqrt{2}
H^{+}\\ H_1+i H_2 \end{array}\right) \,\, , \label{choice}
\end{eqnarray}
and the vacuum expectation values are,
\begin{eqnarray}
<\phi_{1}>=\frac{1}{\sqrt{2}}\left(\begin{array}{c c}
0\\v\end{array}\right) \,  \, ; <\phi_{2}>=0 \,\, .\label{choice2}
\end{eqnarray}
This choice brings the possibility that the SM (new) particles are
collected in the first (second) doublet and the Higgs fields $H_1$
and $H_2$ become the mass eigenstates $h^0$ and $A^0$ respectively
since no mixing occurs between two CP-even neutral bosons $H^0$
and $h^0$ at tree level in this case. As it is seen in eq.
(\ref{lagrangian}), the new Higgs field $\phi_{2}$ is responsible
for the LFV interaction at tree level. With the addition of extra
dimensions, after the compactification on the orbifold $S^1/Z_2$,
the new Higgs field $\phi_{2}$ can be expanded as
\begin{eqnarray}
\phi_{2}(x,y ) & = & {1 \over {\sqrt{2 \pi R}}} \left\{
\phi_{2}^{(0)}(x) + \sqrt{2} \sum_{n=1}^{\infty} \phi_{2}^{(n)}(x)
\cos(ny/R)\right\} \,, \label{SecHiggsField}
\end{eqnarray}
where $\phi_{2}^{(0)}(x)$ ($\phi_{2}^{(n)}(x)$) is  the Higgs
doublet in the four dimensions (the KK modes) including the
charged Higgs boson $H^+$ ($H^{(n)+}$), the neutral CP even-odd
Higgs bosons $h^0$- $A^0$ ($h^{0 (n)}$- $A^{0 (n)}$). The non-zero
$n^{th}$ KK mode of the charged Higgs mass is
$\sqrt{m_{H^\pm}^2+m_n^2}$, and the neutral CP even (odd) Higgs
mass is $\sqrt{m_{h^0}^2+m_n^2}$, ($\sqrt{m_{A^0}^2+m_n^2}$ ),
with the $n$'th level KK particle mass $m_n=n/R$.

The $Z\rightarrow l_1^- l_2^+$ decay exist at least at one loop
level in the 2HDM with the help of the internal neutral Higgs
particles $h^0$ and $A^0$. In Fig. \ref{fig1ver} the necessary
1-loop diagrams, the self energy and vertex diagrams, are given.
With the inclusion of extra dimensions, there exists the
additional contributions due to the KK modes of neutral Higgs
particles. At this stage one needs the lepton-lepton-$S$
($S=h^0,A^0$) interaction, which is modified in the case of the
split fermion scenario and these vertex factors
$V^n_{LR\,(RL)\,ij}$ in the vertices $\bar{\hat{f}}_{iL\,
(R)}\,S^{(n)}(x)\,\cos(ny/R)\, \hat{f}_{j R\, (L)}$,  with the
right (left) handed $i^{th}$ flavor lepton fields $\hat{f}_{j R\,
(L)}$ in five dimensions (see eq. (\ref{gaussianprof})), are
obtained by the integration over the fifth dimension. Finally, the
vertex factor for $n^{th}$ KK mode Higgs fields S read
\begin{eqnarray}
V^n_{LR\,(RL)\,ij}=e^{-n^2\,\sigma^2/4\,R^2}\,e^{-(y_{i L\,
(R)}-y_{j R\, (L)})^2/4 \sigma^2}\, \cos\, [\frac{n\,(y_{i L\,
(R)}+y_{j R\, (L)})}{2\,R}] \,\, . \label{Vij1}
\end{eqnarray}
For $n=0$, this factor becomes $V^0_{LR\,(RL)ij}=e^{-(y_{i L\,
(R)}-y_{j R\, (L)})^2/4 \sigma^2}$ and we define the Yukawa
couplings in four dimensions as
\begin{eqnarray}
\xi^{E}_{ij}\,\Big((\xi^{E \dagger}_{ij})^\dagger\Big)=
V^0_{LR\,(RL)\,ij} \, \xi^{E}_{5\, ij}\,\Big((\xi^{E}_{5\,
ij})^\dagger\Big)/\sqrt{2 \pi R} \,\, . \label{coupl4}
\end{eqnarray}

Now, we would like to present the general effective vertex for the
interaction of on-shell Z-boson with a fermionic current:
\begin{eqnarray}
\Gamma_{\mu}=\gamma_{\mu}(f_V-f_A\ \gamma_5)+
\frac{i}{m_W}\,(f_M+f_E\, \gamma_5)\, \sigma_{\mu\,\nu}\, q^{\nu}
\label{vertex}
\end{eqnarray}
where $q$ is the momentum transfer, $q^2=(p-p')^2$, $f_V$ ($f_A$)
is vector (axial-vector) coupling, $f_M$ ($f_E$) magnetic
(electric) transitions of unlike fermions. Here $p$
($-p^{\prime}$) is the four momentum vector of lepton
(anti-lepton). The  vector (axial-vector) $f_V$ ($f_A$) couplings
and the magnetic (electric) transitions $f_M$ ($f_E$) including
the contributions coming from a single extra dimension can be
obtained as
\begin{eqnarray}
f_V&=&\sum_{i=1}^{3} \Big( f_{iV}^{(0)}+2 \sum_{n=1}^{\infty}
f_{iV}^{(n)} \Big)
\nonumber \, , \\
f_A&=&\sum_{i=1}^{3} \Big( f_{iA}^{(0)}+2 \sum_{n=1}^{\infty}
f_{iA}^{(n)}
\Big)\nonumber \, , \\
f_M&=&\sum_{i=1}^{3} \Big( f_{iM}^{(0)}+2 \sum_{n=1}^{\infty}
f_{iM}^{(n)}
\Big)\nonumber \, , \\
f_E&=&\sum_{i=1}^{3} \Big( f_{iE}^{(0)}+2 \sum_{n=1}^{\infty}
f_{iE}^{(n)} \Big)\, , \label{fVAMEex}
\end{eqnarray}
where $f^{(0)}_{i(V,A,M,E)}$ are the couplings without scalar
boson $S=h^0, A^0$ KK mode contributions and they can be obtained
by taking $n=0$ in eq. (\ref{fVAME}). On the other hand the
couplings $f^{(n)}_{i(V,A,M,E)}$ are the ones due to the KK modes
of the scalar bosons $S=h^0, A^0$ (see eq. (\ref{fVAME})). Here
the summation over the index $i$ represents the sum due to the
internal lepton flavors, namely, $e,\mu,\tau$. We present
$f^{(n)}_{i(V,A,M,E)}$ in the appendix, by taking into account all
the masses of internal leptons and external lepton (anti-lepton).
If we consider two extra dimensions where all the particles are
accessible, the couplings $f^{(n)}_{i(V,A,M,E)}$ appearing in eq.
(\ref{fVAMEex}) should be replaced by $f^{(n,s)}_{i(V,A,M,E)}$ and
they read
\begin{eqnarray}
f_V&=&\sum_{i=1}^{3} \Big( f_{iV}^{(0,0)}+4 \sum_{n,s}^{\infty}
f_{iV}^{(n,s)} \Big)
\nonumber \, , \\
f_A&=&\sum_{i=1}^{3} \Big( f_{iA}^{(0,0)}+4 \sum_{n,s}^{\infty}
f_{iA}^{(n,s)}
\Big)\nonumber \, , \\
f_M&=&\sum_{i=1}^{3} \Big( f_{iM}^{(0,0)}+4 \sum_{n,s}^{\infty}
f_iM^{(n,s)}
\Big)\nonumber \, , \\
f_E&=&\sum_{i=1}^{3} \Big( f_{iE}^{(0,0)}+4 \sum_{n,s}^{\infty}
f_{iE}^{(n,s)} \Big)\, , \label{fVAMEex2}
\end{eqnarray}
where the summation would be done over $n,s=0,1,2 ...$ except
$n=s=0$. (See the appendix for their explicit forms).

Finally, the BR for $Z\rightarrow l_1^-\, l_2^+$ can be written in
terms of the couplings $f_V$, $f_A$, $f_M$ and $f_E$ as
\begin{eqnarray}
BR (Z\rightarrow l_1^-\,l_2^+)=\frac{1}{48\,\pi}\,
\frac{m_Z}{\Gamma_Z}\,
\{|f_V|^2+|f_A|^2+\frac{1}{2\,cos^2\,\theta_W} (|f_M|^2+|f_E|^2)
\} \label{BR1}
\end{eqnarray}
where $\alpha_W=\frac{g^2}{4\,\pi}$ and $\Gamma_Z$ is the total
decay width of Z boson. In our numerical analysis  we consider the
BR due to the production of sum of charged states, namely
\begin{eqnarray}
BR (Z\rightarrow l_1^{\pm}\,l_2^{\pm})= \frac{\Gamma(Z\rightarrow
(\bar{l}_1\,l_2+\bar{l}_2\,l_1)}{\Gamma_Z} \, .\label{BR2}
\end{eqnarray}
%
\section{Discussion}
Since the LFV Z decays $Z\rightarrow l_1^{\pm} l_2^{\pm}$,
$l_1\neq l_2$, exist at least in the one loop level in the 2HDM,
the internal leptons and new scalar bosons drive the interaction
and the physical quantities of these decays are sensitive to the
Yukawa couplings $\bar{\xi}^E_{N,ij}, \, i,j=e, \mu, \tau$, which
are among the free parameters of the model
\footnote{Here, we use the dimensionful coupling
$\bar{\xi}^{E}_{N,ij}$ with the definition
$\xi^{E}_{N,ij}=\sqrt{\frac{4\, G_F}{\sqrt{2}}}\,
\bar{\xi}^{E}_{N,ij}$ where N denotes the word "neutral".}
. Furthermore, respecting the split fermion scenario, which is
based on the idea that the hierarchy of fermion masses are due to
the fermion Gaussian profiles in the extra dimensions, there arise
new parameters, namely, the compactification radius, the fermion
widths and their locations in the new dimesion(s). In this
scenario, the Yukawa couplings in four dimensions appear with a
multiplicative exponential suppression factors after the
integration of the extra dimension (see eq. (\ref{coupl4})). This
factor is coming from the different locations of various flavors
and their left and right handed parts of lepton fields, in the
Yukawa part of the lagrangian. Here, we consider that the
couplings $\bar{\xi}^{E}_{N,ij},\, i,j=e,\mu$ are smaller compared
to $\bar{\xi}^{E}_{N,\tau\, i}\, i=e,\mu,\tau$, respecting the
Sher scenario \cite{Sher}, since latter ones contain heavy
flavors. Furthermore, we assume that, in four dimensions, the
couplings $\bar{\xi}^{E}_{N,ij}$ is symmetric with respect to the
indices $i$ and $j$ and choose the appropriate numerical values
for the Yukawa couplings, by respecting the current experimental
measurements. The upper limit of $\bar{\xi}^{E}_{N,\tau \mu}$ is
predicted as $30\, GeV$ (see \cite{Iltananomuon} and references
therein) by using the experimental uncertainty, $10^{-9}$, in the
measurement of the muon anomalous magnetic moment and assuming
that the new physics effects can not exceed this uncertainty.
Using this upper limit and the experimental upper bound of BR of
$\mu\rightarrow e \gamma$ decay, BR $\leq 1.2\times 10^{-11}$
\cite{Brooks}, the coupling $\bar{\xi}^{E}_{N,\tau e}$ can be
restricted in the range, $10^{-3}-10^{-2}\, GeV$ (see
\cite{Iltan1}). For the Yukawa coupling $\bar{\xi}^{E}_{N,\tau
\tau}$, we have no explicit restriction region and we use the
numerical values which are greater than $\bar{\xi}^{E}_{N,\tau
\mu}$.

Our study is devoted to the prediction of the effects of the extra
dimensions on the BR of the LFV processes $Z\rightarrow l_1^{\pm}
l_2^{\pm}$, in the split fermion scenario, in the framework of the
2HDM. The compactification scale $1/R$, which is one of the free
parameter of the model, should be restricted. In the literature,
there exist numerous constraints for this scale, in the case of
the single extra dimension, in the split fermion scenario:
\begin{itemize}
\item $1/R> 800\,\, GeV$ due to the direct limits from searching
for KK gauge bosons.
\item  $1/R> 1.0 \,\, TeV$ from $B\rightarrow \phi \, K_S$, $1/R >
500\,\, GeV$ from $B\rightarrow \psi \, K_S$ and $1/R
> 800\,\, GeV$ from the upper limit of the $BR$, $BR \, (B_s
\rightarrow \mu^+ \mu^-)< 2.6\,\times 10^{-6}$ \cite{Hewett}.
\item A far more stringent limit $1/R> 3.0\,\, TeV$ \cite{Rizzo}
coming from the precision electro weak bounds on higher
dimensional operators generated by KK exchange
\end{itemize}
In our numerical analysis, we choose an appropriate range for the
compactification scale $1/R$, by respecting these limits in the
case of a single extra dimension. For two extra dimensions we used
the same broad range for $1/R$.
Throughout our calculations we use the input values given in Table
(\ref{input}).
\begin{table}[h]
        \begin{center}
        \begin{tabular}{|l|l|}
        \hline
        \multicolumn{1}{|c|}{Parameter} &
                \multicolumn{1}{|c|}{Value}     \\
        \hline \hline
        $m_{\mu}$                   & $0.106$ (GeV) \\
        $m_{\tau}$                   & $1.78$ (GeV) \\
        $m_{W}$             & $80.26$ (GeV) \\
        $m_{Z}$             & $91.19$ (GeV) \\
        $m_{h^0}$             & $100$ (GeV) \\
        $m_{A^0}$             & $200$ (GeV) \\
        $G_F$             & $1.16637 10^{-5} (GeV^{-2})$  \\
        $\Gamma_Z$                  & $2.490\, (GeV)$  \\
        $sin\,\theta_W$               & $\sqrt{0.2325}$ \\
        \hline
        \end{tabular}
        \end{center}
\caption{The values of the input parameters used in the numerical
          calculations.}
\label{input}
\end{table}

In the present work, we estimate the BRs of the LFV Z decays in a
single and two extra dimensions, by considering split leptons with
a possible set of locations. For a single extra dimension (two
extra dimensions) we use the estimated location of the leptons
given in eq. (\ref{location}) (eq. (\ref{location2})) to calculate
the lepton-lepton-Higgs scalar KK mode vertices. In the case of
two extra dimensions we study two possibilities: the leptons are
restricted to the fifth extra dimension, with non-zero Gaussian
profiles and the leptons have non-zero Gaussian profiles also in
the sixth dimension. In the former one, the enhancements in the
BRs of the present decays are relatively large due to the well
known KK mode abundance of Higgs fields. However, in the latter
the additional exponential factor appearing in the second
summation further suppresses the BRs.

Fig. \ref{ZmueRr} is devoted to the compactification scale $1/R$
dependence of the BR $\,(Z\rightarrow \mu^{\pm}\, e^{\pm})$ for
$\bar{\xi}^{D}_{N,\tau e}=0.01\, GeV$, $\bar{\xi}^{D}_{N,\tau
\mu}=1\,GeV$ and $\rho=\sigma/R=0.01$. Here the solid (dashed,
small dashed, dotted) line represents the BR without extra
dimension (with a single extra dimension, with two extra
dimensions where the leptons have non-zero Gaussian profiles in
the fifth extra dimension, with two extra dimensions where the
leptons have non-zero Gaussian profiles in both extra dimensions).
It is observed that BR is at the order of the magnitude of
$10^{-14}$ without extra dimensions and it is the  weakly
sensitive to the parameter $1/R$, for the $1/R>500\, GeV$, for a
single extra dimension. In the case of two extra dimensions, the
enhancement of the BR is relatively larger, near one order of
magnitude, for $1/R\sim 0.6\, TeV$, due to the Higgs scalar KK
mode abundances. However, these contributions do not increase
extremely due to the suppression exponential factor appearing in
the summations. The enhancement in the BR becomes weak for
$1/R>3.0\, TeV$. Furthermore, the numerical values of BRs are
slightly greater in the case that the leptons have non-zero
Gaussian profiles only in the fifth extra dimension.

In Fig. \ref{ZtaueRr}, we present the compactification scale $1/R$
dependence of the BR $\,(Z\rightarrow \tau^{\pm}\, e^{\pm})$ for
$\bar{\xi}^{D}_{N,\tau e}=0.01\, GeV$, $\bar{\xi}^{D}_{N,\tau
\tau}=10\,GeV$ and $\rho=0.01$. Here the solid (dashed, small
dashed, dotted) line represents the BR without extra dimension
(with a single extra dimension, with two extra dimensions where
the leptons have non-zero Gaussian profiles in the fifth extra
dimension, with two extra dimensions where the leptons have
non-zero Gaussian profiles in both extra dimensions). This figure
shows that the BR is at the order of the magnitude of $10^{-12}$
without extra dimensions. Similar to the previous decay, the BR is
weakly sensitive to the parameter $1/R$, for the $1/R>500\, GeV$,
for a single extra dimension. In the case of two extra dimensions,
the enhancement of the BR is almost one order larger than the one
with a single extra dimension, for $1/R\sim 0.6\, TeV$ and this
enhancement becomes weak for $1/R>3.0\, TeV$.

Fig. \ref{ZtaumuRr} represents the compactification scale $1/R$
dependence of the BR $\,(Z\rightarrow \tau^{\pm}\, \mu^{\pm})$ for
$\bar{\xi}^{D}_{N,\tau \mu}=1\, GeV$, $\bar{\xi}^{D}_{N,\tau
\tau}=10\,GeV$ and $\rho=0.01$. Here the solid (dashed, small
dashed, dotted) line represents the BR without extra dimension
(with a single extra dimension, with two extra dimensions where
the leptons have non-zero Gaussian profiles in the fifth extra
dimension, with two extra dimensions where the leptons have
non-zero Gaussian profiles in both extra dimensions). For this
decay, the BR is observed at the order of the magnitude of
$10^{-8}$ without extra dimensions. The BR is weakly sensitive to
the parameter $1/R$, for the $1/R>500\, GeV$, for a single extra
dimension. In the case of two extra dimensions, the enhancement of
the BR is more than one order larger than the one with a single
extra dimension, for $1/R\sim 0.6\, TeV$ and this enhancement
becomes also weak for $1/R>3.0\, TeV$.

Now we would like to estimate the sensitivity of the BRs of the Z
decays under consideration to the Gaussian widths, $\sigma=\rho\,
R$ of leptons, where $\rho$ is the free parameter which regulates
the amount of width in the extra dimension.

Fig. \ref{Zmuero} (\ref{Ztauero} \textbf{;} \ref{Ztaumuro}) shows
the parameter $\rho$ dependence of the BR of the decay
$Z\rightarrow \mu^{\pm}\, e^{\pm}$ ($\,(Z\rightarrow \tau^{\pm}\,
e^{\pm})$ \textbf{;} $\,(Z\rightarrow \tau^{\pm}\, \mu^{\pm})$)
for $1/R=500\, GeV$,  and the real couplings
$\bar{\xi}^{E}_{N,\tau \mu} =1\, GeV$, $\bar{\xi}^{E}_{N,\tau e}
=0.01\, GeV$ ($\bar{\xi}^{E}_{N,\tau \tau} =10\, GeV$,
$\bar{\xi}^{E}_{N,\tau e} =0.01\, GeV$ \textbf{;}
$\bar{\xi}^{E}_{N,\tau \tau} =10\, GeV$, $\bar{\xi}^{E}_{N,\tau
\mu} =1\, GeV$). Here the solid (dashed, small dashed, dotted)
line represents the BR without extra dimension (with a single
extra dimension, with two extra dimensions where the leptons have
non-zero Gaussian profiles in the fifth extra dimension, with two
extra dimensions where the leptons have non-zero Gaussian profiles
in both extra dimensions). It is observed that the BR of the decay
$Z\rightarrow \mu^{\pm}\, e^{\pm}$ ($\,(Z\rightarrow \tau^{\pm}\,
e^{\pm})$ \textbf{;} $\,(Z\rightarrow \tau^{\pm}\, \mu^{\pm})$)
increases almost $50\%$ ($50\%$ \textbf{;} $50\%$) for  $\rho >
0.03$ and reaches $70\%$ ($74\%$ \textbf{;} $74\%$) for  $\rho
\sim 0.001$, in the case of a single extra dimension. This shows
that the sensitivities of the BRs of the present decays to the
parameter $\rho$ are not weak, especially for its small values.
For two extra dimensions, these sensitivities increases
considerably. For the decay $\,Z\rightarrow \mu^{\pm}\, e^{\pm}$
($\,(Z\rightarrow \tau^{\pm}\, e^{\pm})$ \textbf{;}
$\,(Z\rightarrow \tau^{\pm}\, \mu^{\pm})$), the BR enhances more
than one order ( more than one order \textbf{;} $\sim 2000\%$) for
the intermediate values of the parameter $\rho$, in the case of
two extra dimensions where the leptons have non-zero Gaussian
profiles in the fifth extra dimension. This enhancement reaches
$\sim 4000\%$  ($\sim 4000\%$  \textbf{;} $\sim 4000\%$) for $\rho
\sim 0.001$. In the case that the leptons have non-zero Gaussian
profiles in both extra dimensions, the BR enhances less than one
order (less than one order \textbf{;} more than one order), for
the intermediate values of the parameter $\rho$, for the decay
$\,Z\rightarrow \mu^{\pm}\, e^{\pm}$ ($\,(Z\rightarrow
\tau^{\pm}\, e^{\pm})$ \textbf{;} $\,(Z\rightarrow \tau^{\pm}\,
\mu^{\pm})$). It is shown that the sensitivities of the BRs of
studied LFV decays are considerably large for two extra dimensions
and the BRs enhances more than one order, especially for the LFV Z
decays with heavy lepton flavors.

As a summary, the BR is weakly sensitive to the parameter $1/R$
for $1/R>500\, GeV$ for a single extra dimension, however, there
is an enhancement in the BR, even more than one order for the
scale $1/R \sim 600\, GeV$. This enhancement decreases with the
increasing values of the scale $1/R$. Furthermore, the BR is
sensitive to the parameter $\rho$ especially for two extra
dimensions case. Therefore, the LFV Z decays are worthwhile to
study and with the help of the forthcoming more accurate
experimental measurements of the these decays, especially the
$\,Z\rightarrow \tau^{\pm}\, \mu^{\pm}$ one, the valuable
information can be obtained to detect the effects due to the extra
dimensions in the case of split fermion scenario.
\section{Acknowledgement}
This work has been supported by the Turkish Academy of Sciences in
the framework of the Young Scientist Award Program.
(EOI-TUBA-GEBIP/2001-1-8)
%
%
\section{The explicit expressions appearing in the text}
Here we present the explicit expressions for $f_{iV}^{(n)}$,
$f_{iA}^{(n)}$, $f_{iM}^{(n)}$ and $f_{iE}^{(n)}$ \cite{EiltZl1l2}
(see eq. (\ref{fVAMEex})):
\begin{eqnarray}
f_{iV}^{(n)}&=& \frac{g}{64\,\pi^2\,cos\,\theta_W} \int_0^1\, dx
\, \frac{1}{m^2_{l_2^+}-m^2_{l_1^-}} \Bigg \{ c_V \,
(m_{l_2^+}+m_{l_1^-})
\nonumber \\
&\Bigg(& (-m_i \, \eta^+_i + m_{l_1^-} (-1+x)\, \eta_i^V)\, ln \,
\frac{L^{self}_ {1,\,h^0}}{\mu^2}+ (m_i \, \eta^+_i - m_{l_2^+}
(-1+x)\, \eta_i^V)\, ln \, \frac{L^{self}_{2,\, h^0}}{\mu^2}
\nonumber \\ &+& (m_i \, \eta^+_i + m_{l_1^-} (-1+x)\, \eta_i^V)\,
ln \, \frac{L^{self}_{1,\, A^0}}{\mu^2} - (m_i \, \eta^+_i +
m_{l_2^+} (-1+x) \,\eta_i^V)\, ln \, \frac{L^{self}_{2,\,
A^0}}{\mu^2} \Bigg) \nonumber \\ &+&
c_A \, (m_{l_2^+}-m_{l_1^-}) \nonumber \\
&\Bigg ( & (-m_i \, \eta^-_i + m_{l_1^-} (-1+x)\, \eta_i^A)\, ln
\, \frac{L^{self}_{1,\, h^0}}{\mu^2} + (m_i \, \eta^-_i +
m_{l_2^+} (-1+x)\, \eta_i^A)\, ln \, \frac{L^{self}_{2,\,
h^0}}{\mu^2} \nonumber \\ &+& (m_i \, \eta^-_i + m_{l_1^-}
(-1+x)\, \eta_i^A)\, ln \, \frac{L^{self}_{1,\, A^0}}{\mu^2} +
(-m_i \, \eta^-_i + m_{l_2^+} (-1+x)\, \eta_i^A)\, ln \,
\frac{L^{self}_{2,\, A^0}}{\mu^2} \Bigg) \Bigg \} \nonumber \\ &-&
\frac{g}{64\,\pi^2\,cos\,\theta_W} \int_0^1\,dx\, \int_0^{1-x} \,
dy \, \Bigg \{ m_i^2 \,(c_A\,
\eta_i^A-c_V\,\eta_i^V)\,(\frac{1}{L^{ver}_{A^0}}+
\frac{1}{L^{ver}_{h^0}}) \nonumber \\ &-& (1-x-y)\,m_i\, \Bigg(
c_A\,  (m_{l_2^+}-m_{l_1^-})\, \eta_i^- \,(\frac{1}{L^{ver}_{h^0}}
- \frac{1}{L^{ver}_{A^0}})+ c_V\, (m_{l_2^+}+m_{l_1^-})\, \eta_i^+
\, (\frac{1}{L^{ver}_{h^0}} + \frac{1}{L^{ver}_{A^0}}) \Bigg)
\nonumber \\ &-& (c_A\, \eta_i^A+c_V\,\eta_i^V) \Bigg (
-2+(q^2\,x\,y+m_{l_1^-}\,m_{l_2^+}\, (-1+x+y)^2)\,
(\frac{1}{L^{ver}_{h^0}} +
\frac{1}{L^{ver}_{A^0}})-ln\,\frac{L^{ver}_{h^0}}{\mu^2}\,
\frac{L^{ver}_{A^0}}{\mu^2} \Bigg ) \nonumber \\ &-&
(m_{l_2^+}+m_{l_1^-})\, (1-x-y)\, \Bigg (
\frac{\eta_i^A\,(x\,m_{l_1^-} +y\,m_{l_2^+})+m_i\,\eta_i^-}
{2\,L^{ver}_{A^0\,h^0}}+\frac{\eta_i^A\,(x\,m_{l_1^-}
+y\,m_{l_2^+})- m_i\,\eta_i^-}{2\,L^{ver}_{h^0\,A^0}} \Bigg )
\nonumber \\ &+& \frac{1}{2}\eta_i^A\,
ln\,\frac{L^{ver}_{A^0\,h^0}}{\mu^2}\,
\frac{L^{ver}_{h^0\,A^0}}{\mu^2}
\Bigg \}\,, \nonumber \\
f_{iA}^{(n)}&=& \frac{-g}{64\,\pi^2\,cos\,\theta_W} \int_0^1\, dx
\, \frac{1}{m^2_{l_2^+}-m^2_{l_1^-}} \Bigg \{ c_V \,
(m_{l_2^+}-m_{l_1^-})
\nonumber \\
&\Bigg(& (m_i \, \eta^-_i + m_{l_1^-} (-1+x)\, \eta_i^A)\, ln \,
\frac{L^{self}_{1,\,A^0}}{\mu^2} + (-m_i \, \eta^-_i + m_{l_2^+}
(-1+x)\, \eta_i^A)\, ln \, \frac{L^{self}_ {2,\,A^0}}{\mu^2}
\nonumber \\ &+& (-m_i \, \eta^-_i + m_{l_1^-} (-1+x)\,
\eta_i^A)\, ln \, \frac{L^{self}_{1,\, h^0}}{\mu^2}+ (m_i \,
\eta^-_i + m_{l_2^+} (-1+x)\, \eta_i^A)\, ln \,
\frac{L^{self}_{2,\,h^0}}{\mu^2} \Bigg) \nonumber \\ &+&
c_A \, (m_{l_2^+}+m_{l_1^-}) \nonumber \\
&\Bigg(& (m_i \, \eta^+_i + m_{l_1^-} (-1+x)\, \eta_i^V)\, ln \,
\frac{L^{self}_{1,\, A^0}}{\mu^2}- (m_i \, \eta^+_i + m_{l_2^+}
(-1+x)\, \eta_i^V)\, ln \, \frac{L^{self}_{2,\,A^0}}{\mu^2}
\nonumber \\ &+& (-m_i \, \eta^+_i + m_{l_1^-} (-1+x)\,
\eta_i^V)\, ln \, \frac{L^{self}_{1,\, h^0}}{\mu^2} + (m_i \,
\eta^+_i - m_{l_2^+} (-1+x)\, \eta_i^V)\, \frac{ln \,
L^{self}_{2,\,h^0}}{\mu^2} \Bigg) \Bigg \} \nonumber \\ &+&
\frac{g}{64\,\pi^2\,cos\,\theta_W} \int_0^1\,dx\, \int_0^{1-x} \,
dy \, \Bigg \{ m_i^2 \,(c_V\,
\eta_i^A-c_A\,\eta_i^V)\,(\frac{1}{L^{ver}_{A^0}}+
\frac{1}{L^{ver}_{h^0}}) \nonumber \\ &-& m_i\, (1-x-y)\, \Bigg(
c_V\, (m_{l_2^+}-m_{l_1^-}) \,\eta_i^- + c_A\,
(m_{l_2^+}+m_{l_1^-})\, \eta_i^+ \Bigg) \,(\frac{1}
{L^{ver}_{h^0}} - \frac{1}{L^{ver}_{A^0}}) \nonumber \\ &+& (c_V\,
\eta_i^A+c_A\,\eta_i^V) \Bigg(-2+(q^2\,x\,y-m_{l_1^-}\,m_{l_2^+}\,
(-1+x+y)^2) (\frac{1}{L^{ver}_{h^0}}+\frac{1}{L^{ver}_{A^0}})-
ln\,\frac{L^{ver}_{h^0}}{\mu^2}\,\frac{L^{ver}_{A^0}}{\mu^2}
\Bigg) \nonumber \\ &-& (m_{l_2^+}-m_{l_1^-})\, (1-x-y)\, \Bigg(
\frac{\eta_i^V\,(x\,m_{l_1^-} -y\,m_{l_2^+})+m_i\,\eta_i^+}
{2\,L^{ver}_{A^0\,h^0}}+ \frac{\eta_i^V\,(x\,m_{l_1^-}
-y\,m_{l_2^+})-m_i\, \eta_i^+}{2\,L^{ver}_{h^0\,A^0}}
\Bigg)\nonumber \\
&-& \frac{1}{2} \eta_i^V\, ln\,\frac{L^{ver}_{A^0\,h^0}}{\mu^2}\,
\frac{L^{ver}_{h^0\,A^0}}{\mu^2}
\Bigg \} \nonumber \,,\\
f_{iM}^{(n)}&=&-\frac{g\, m_W}{64\,\pi^2\,cos\,\theta_W}
\int_0^1\,dx\, \int_0^{1-x} \, dy \, \Bigg \{ \Bigg(
(1-x-y)\,(c_V\, \eta_i^V+c_A\,\eta_i^A)\, (x\,m_{l_1^-}
+y\,m_{l_2^+}) \nonumber
\\ &+& \, m_i\,(c_A\, (x-y)\,\eta_i^-+c_V\,\eta_i^+\,(x+y))\Bigg )
\,\frac{1}{L^{ver}_{h^0}} \nonumber \\ &+& \Bigg( (1-x-y)\, (c_V\,
\eta_i^V+c_A\,\eta_i^A)\, (x\,m_{l_1^-} +y\,m_{l_2^+})
-m_i\,(c_A\, (x-y)\,\eta_i^-+c_V\,\eta_i^+\,(x+y))\Bigg )
\,\frac{1}{L^{ver}_{A^0}} \nonumber \\ &-& (1-x-y) \Bigg
(\frac{\eta_i^A\,(x\,m_{l_1^-} +y\,m_{l_2^+})}{2}\, \Big (
\frac{1}{L^{ver}_{A^0\,h^0}}+\frac{1}{L^{ver}_{h^0\,A^0}} \Big )
+\frac{m_i\,\eta_i^-} {2} \, \Big ( \frac{1}{L^{ver}_{h^0\,A^0}}-
\frac{1}{L^{ver}_{A^0\,h^0}} \Big ) \Bigg ) \Bigg \} \,,\nonumber \\
f_{iE}^{(n)}&=&-\frac{g\, m_W}{64\,\pi^2\, cos\,\theta_W}
\int_0^1\,dx\, \int_0^{1-x} \, dy \, \Bigg \{ \Bigg( (1-x-y)\,\Big
( -(c_V\, \eta_i^A+c_A\,\eta_i^V)\, (x\,m_{l_1^-} -y\, m_{l_2^+})
\Big) \nonumber \\ &-& m_i\, (c_A\,
(x-y)\,\eta_i^++c_V\,\eta_i^-\,(x+y))\Bigg )\,
\frac{1}{L^{ver}_{h^0}} \nonumber \\ &+& \Bigg ( (1-x-y)\,\Big (
-(c_V\, \eta_i^A+c_A\,\eta_i^V)\, (x\,m_{l_1^-} - y\, m_{l_2^+})
\Big ) + m_i\,(c_A\, (x-y)\,\eta_i^++c_V\,\eta_i^-\,(x+y)) \Bigg )
\,\frac{1}{L^{ver}_{A^0}} \nonumber \\&+& (1-x-y)\, \Bigg (
\frac{\eta_i^V}{2}\,(m_{l_1^-}\,x-m_{l_2^+}\, y)\, \, \Big (
\frac{1}{L^{ver}_{A^0\,h^0}}+\frac{1}{L^{ver}_{h^0\,A^0}} \Big )
+\frac{m_i\,\eta_i^+}{2}\, \Big (
\frac{1}{L^{ver}_{A^0\,h^0}}-\frac{1}{L^{ver}_{h^0\,A^0}} \Big )
\Bigg ) \Bigg \}, \label{fVAME}
\end{eqnarray}
where
\begin{eqnarray}
L^{self}_{1,\,h^0}&=& m^{(n)
2}_{h^0}\,(1-x)+(m_i^2-m^2_{l_1^-}\,(1-x))\,x
\nonumber \, , \\
L^{self}_{1,\,A^0}&=&L^{self}_{1,\,h^0}(m^{(n)}_{h^0}\rightarrow
m^{(n)}_{A^0})
\nonumber \, , \\
L^{self}_{2,\,h^0}&=&L^{self}_{1,\,h^0}(m_{l_1^-}\rightarrow
m_{l_2^+})
\nonumber \, , \\
L^{self}_{2,\,A^0}&=&L^{self}_{1,\,A^0}(m_{l_1^-}\rightarrow
m_{l_2^+})
\nonumber \, , \\
L^{ver}_{h^0}&=& m^{(n) 2}_{h^0}\,(1-x-y)+m_i^2\,(x+y)-q^2\,x\,y
\nonumber \, , \\
L^{ver}_{h^0\,A^0}&=&m^{(n) 2}_{A^0}\,x+m_i^2\,(1-x-y)+(m^{(n)
2}_{h^0}-q^2\,x)\,y
\nonumber \, , \\
L^{ver}_{A^0}&=&L^{ver}_{h^0}(m^{(n)}_{h^0}\rightarrow
m^{(n)}_{A^0})
\nonumber \, , \\
L^{ver}_{A^0\,h^0}&=&L^{ver}_{h^0\,A^0}(m^{(n)}_{h^0}\rightarrow
m^{(n)}_{A^0}) \, , \label{Lh0A0}
\end{eqnarray}
and
\begin{eqnarray}
\eta_i^V&=& e^{-n^2\,\sigma^2/2\,R^2}\,\{ c_n \,(l_1,i)\,c_n
\,(l_2,i) \,\xi^{E}_{il_1}\xi^{E\,*}_{il_2}+ c'_n \,(l_1,i)\,c'_n
\,(l_2,i)\, \xi^{E\,*}_{l_1i} \xi^{E}_{l_2 i} \}\nonumber \, , \\
\eta_i^A&=&e^{-n^2\,\sigma^2/2\,R^2}\,\{ c_n \,(l_1,i)\,c_n
\,(l_2,i) \,\xi^{E}_{il_1}\xi^{E\,*}_{il_2} -c'_n \,(l_1,i)\,c'_n
\,(l_2,i) \,
\xi^{E\,*}_{l_1i} \xi^{E}_{l_2 i}\} \nonumber \, , \\
\eta_i^+&=&e^{-n^2\,\sigma^2/2\,R^2}\,\{ c'_n \,(l_1,i)\,c_n
\,(l_2,i) \, \xi^{E\,*}_{l_1i}\xi^{E\,*}_{il_2}+ c_n
\,(l_1,i)\,c'_n \,(l_2,i) \,
\xi^{E}_{il_1} \xi^{E}_{l_2 i} \,\} \nonumber \, , \\
\eta_i^-&=&e^{-n^2\,\sigma^2/2\,R^2}\,\{ c'_n \,(l_1,i)\,c'_n
\,(l_2,i) \, \xi^{E\,*}_{l_1i}\xi^{E\,*}_{il_2}- c_n
\,(l_1,i)\,c'_n \,(l_2,i) \, \xi^{E}_{il_1} \xi^{E}_{l_2 i}\}\, .
\label{etaVA}
\end{eqnarray}
The parameters $c_V$ and $c_A$ are $c_A=-\frac{1}{4}$ and
$c_V=\frac{1}{4}-sin^2\,\theta_W$ and the masses $m^{(n)}_{S}$
read $m^{(n)}_{S}=\sqrt{m_S^2+n^2/R^2}$, where $R$ is the
compactification radius. In eq. (\ref{etaVA}) the flavor changing
couplings $\xi ^{E}_{i l_j}$ represent the effective interaction
between the internal lepton $i$, ($i=e,\mu,\tau$) and outgoing
(incoming) $j=1\,(j=2)$ one. The parameters $c_n\,(f,i)$,
$c'_n\,(f,i)$ read
\begin{eqnarray}
c_n \,(f,i)&=&\cos[\frac{n\,(y_{f R}+y_{i L})}{2\, R}]\,
\,, \nonumber \\
c'_n \,(f,i)&=&\cos[\frac{n\,(y_{f L}+y_{i R})}{2\, R}]\, .
\label{coeff}
\end{eqnarray}
In the case of two extra dimensions which all the particles feel,
the parameters $c_n\,(f,i)$ and $c'_n\,(f,i)$ are replaced by
\begin{eqnarray}
c_{(n,s)}\, (f,i)&=&\cos[\frac{n\,(y_{f R}+y_{i L})+ s\,(z_{f
R}+z_{i L})}{2\, R}]\, \,,
\nonumber \\
c'_{(n,s)}(f,i)&=&\cos[\frac{n\,(y_{f L}+y_{i R}) + s\,(z_{f
L}+z_{i R})}{2\, R}]\, , \label{coeff22}
\end{eqnarray}
and the exponential factor $e^{-n^2\,\sigma^2/2\,R^2}$ becomes
$e^{-(n^2+s^2)\,\sigma^2/2\,R^2}$. Furthermore, the masses
$m^{(n)}_{S}$ are replaced by $m^{(n,s)}_{S}$,
$m^{(n,s)}_{S}=\sqrt{m^2_{S}+m_n^2+m_s^2}$, with $m_n=n/R, \,
m_s=s/R$. Here we use a possible positions of left handed and
right handed leptons in the two extra dimensions, by using the
observed masses \footnote{The calculation is similar to the one
presented in \cite{Mirabelli} which is done for a single extra
dimension.}. With the assumption that the lepton mass matrix is
diagonal, one of the possible set of locations for the Gaussian
peaks of the lepton fields in the two extra dimensions can be
obtained as \cite{IltanEDMSplit}
\begin{eqnarray}
P_{l_i}=\sqrt{2}\,\sigma\, \left(\begin{array}{c c c}
(8.417,8.417)\\(1.0,1.0)\\(0.0,0.0)
\end{array}\right)\,,\,\,\,\,
P_{e_i}=\sqrt{2}\,\sigma\, \left(\begin{array} {c c c}
(4.7913,4.7913)\\(3.7913,3.7913)\\(-2.2272,-2.2272)
\end{array}\right)
 \,\, , \label{location2}
\end{eqnarray}
where the numbers in the parenthesis denote the y and z
coordinates of the location of the Gaussian peaks of lepton
flavors in the extra dimensions. Here we choose the same numbers
for the y and z locations of the Gaussian peaks.

Finally, the couplings $\xi^{E}_{l_ji}$ may be complex in general
and they can be parametrized as
\begin{eqnarray}
\xi^{E}_{l_ij }=|\xi^{E}_{l_ij }|\, e^{i \theta_{ij}} \,\, ,
\label{xi}
\end{eqnarray}
where $i,l_j$ denote the lepton flavors and $\theta_{ij}$ are CP
violating parameters which are the possible sources of the lepton
EDM. However, in the present work we take these couplings real.
\newpage
\newpage
\begin{figure}[htb]
\vskip 0.0truein \centering \epsfxsize=6.0in
\leavevmode\epsffile{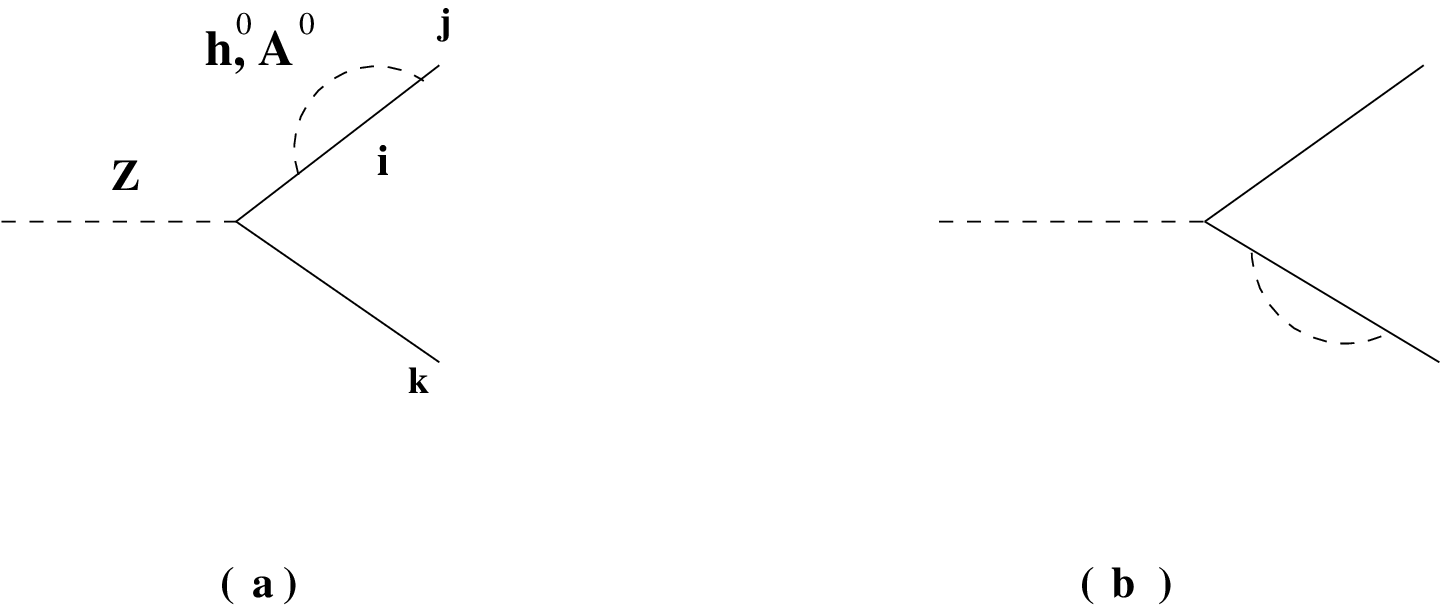} \vskip 0.5truein
\end{figure}
\begin{figure}[htb]
\vskip -0.5truein \centering \epsfxsize=6.0in
\leavevmode\epsffile{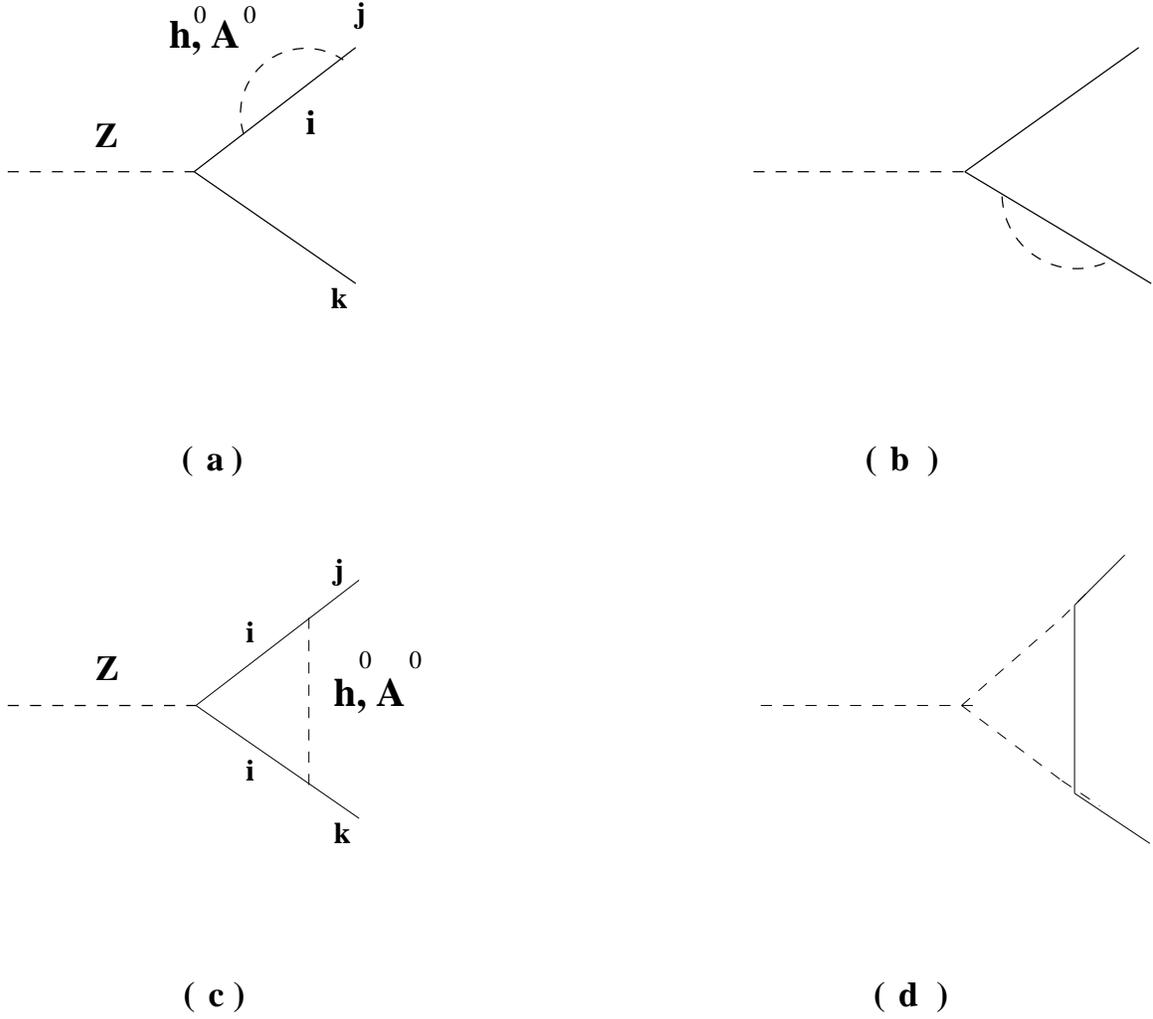} \vskip 0.5truein \caption[]{One
loop diagrams contribute to $Z\rightarrow k^+\,j^-$ decay due to
the neutral Higgs bosons $h_0$ and $A_0$ in the 2HDM. $i$
represents the internal, $j$ ($k$) outgoing (incoming) lepton,
dashed lines the vector field Z, $h_0$ and $A_0$ fields. In the
case 5 (6) dimensions there exits also the KK modes of $h_0$ and
$A_0$ fields.} \label{fig1ver}
\end{figure}
\newpage
\begin{figure}[htb]
\vskip -3.0truein \centering \epsfxsize=6.8in
\leavevmode\epsffile{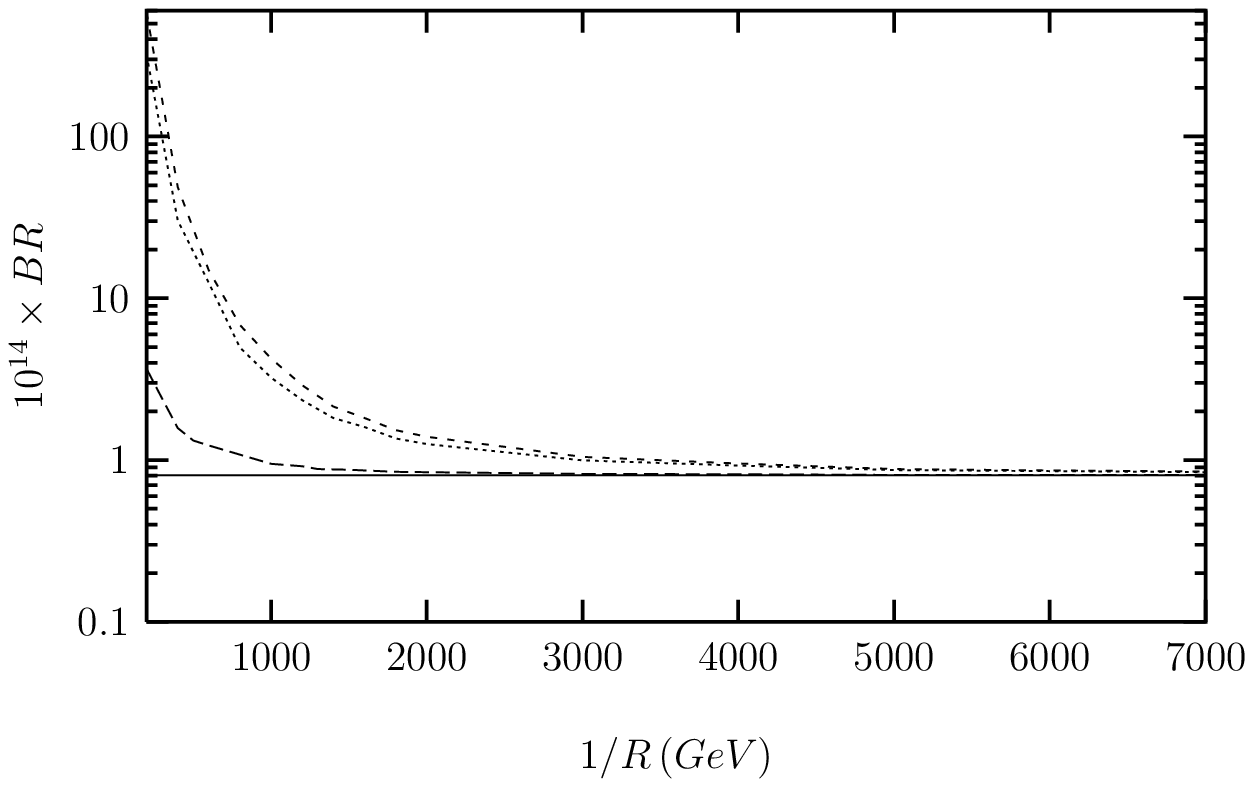} \vskip -3.0truein
\caption[]{BR($\,(Z\rightarrow \mu^{\pm}\, e^{\pm})$) with respect
to the scale $1/R$ for $\rho=0.01$, $\bar{\xi}^{E}_{N,\tau
e}=0.01\, GeV$, $\bar{\xi}^{E}_{N,\tau \mu}=1\,GeV$. Here the
solid (dashed, small dashed, dotted) line represents the BR
without extra dimension (with a single extra dimension, with two
extra dimensions where the leptons have non-zero Gaussian profiles
in the fifth extra dimension, with two extra dimensions where the
leptons have non-zero Gaussian profiles in both extra dimensions)}
\label{ZmueRr}
\end{figure}
\begin{figure}[htb]
\vskip -3.0truein \centering \epsfxsize=6.8in
\leavevmode\epsffile{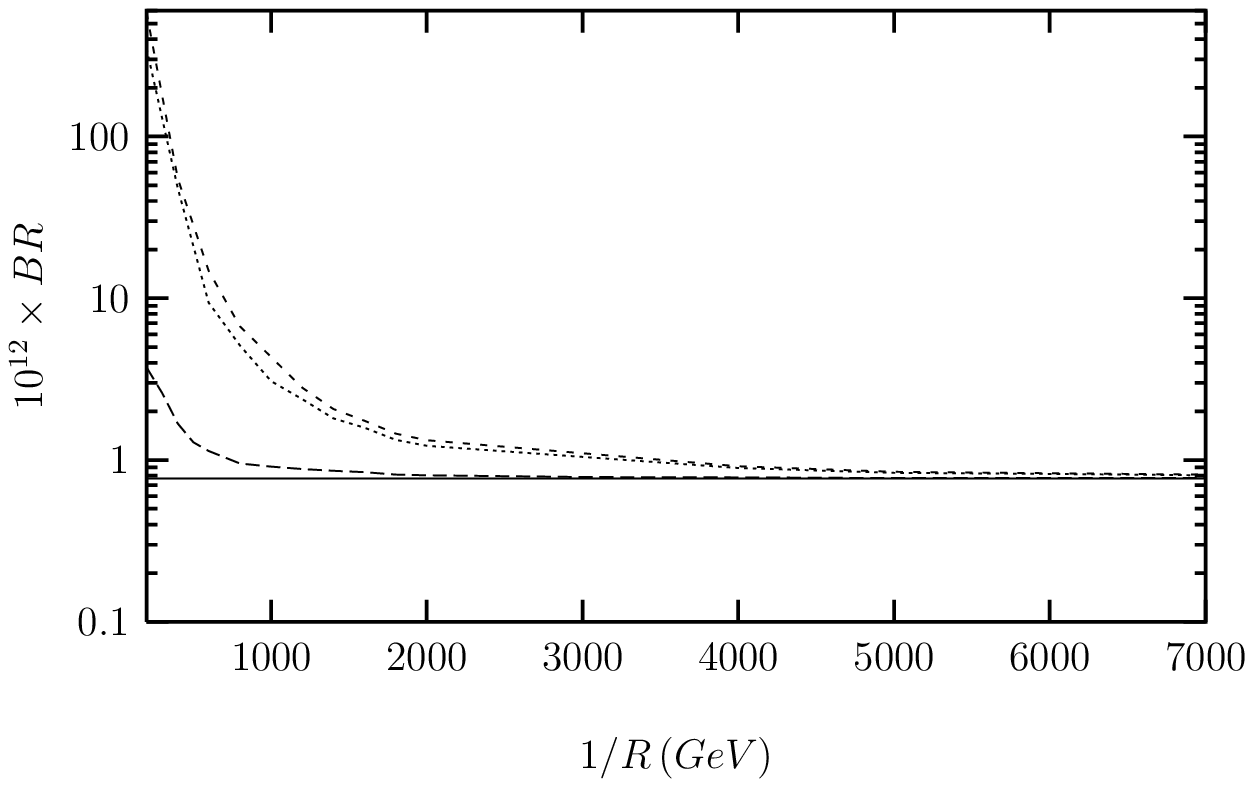} \vskip -3.0truein
\caption[]{BR($\,(Z\rightarrow \tau^{\pm}\, e^{\pm})$) with
respect to the scale $1/R$ for $\rho=0.01$, $\bar{\xi}^{E}_{N,\tau
e}=0.01\, GeV$, $\bar{\xi}^{E}_{N,\tau \tau}=10\,GeV$. Here the
solid (dashed, small dashed, dotted) line represents the BR
without extra dimension (with a single extra dimension, with two
extra dimensions where the leptons have non-zero Gaussian profiles
in the fifth extra dimension, with two extra dimensions where the
leptons have non-zero Gaussian profiles in both extra dimensions)}
\label{ZtaueRr}
\end{figure}
\begin{figure}[htb]
\vskip -3.0truein \centering \epsfxsize=6.8in
\leavevmode\epsffile{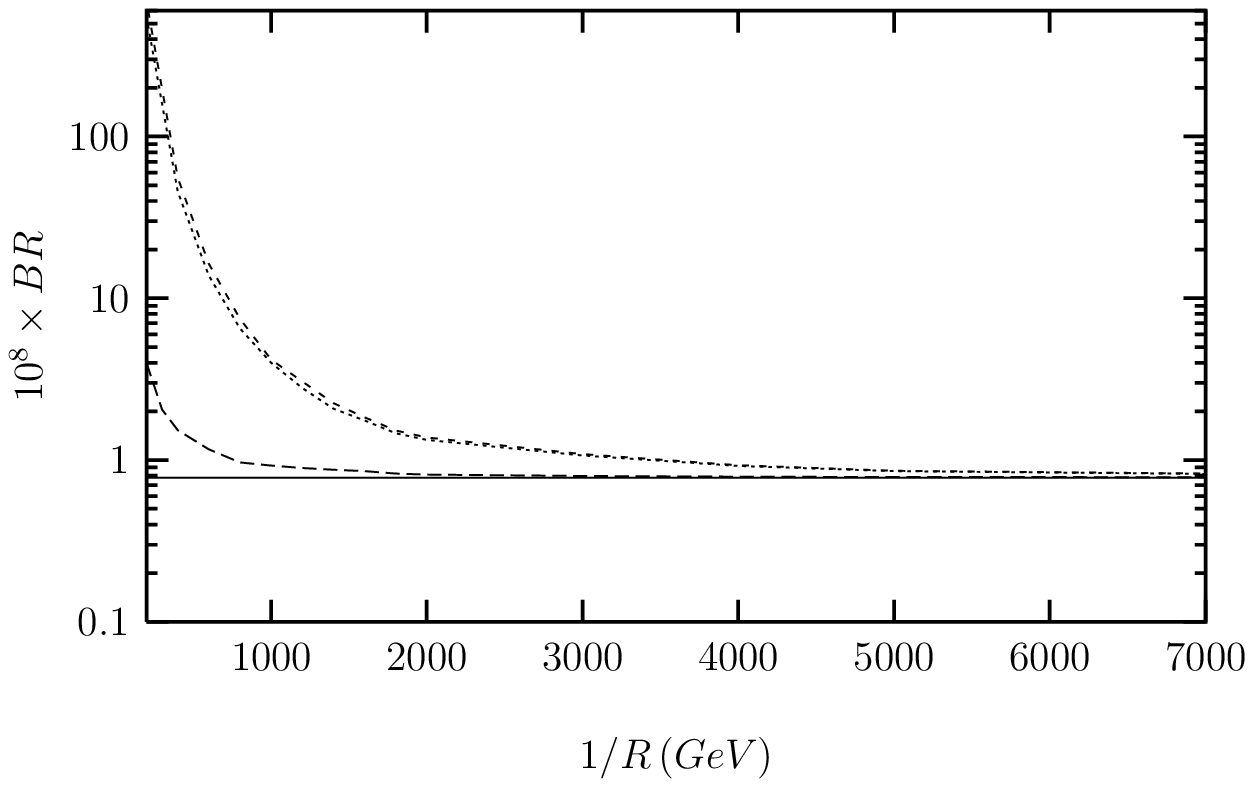} \vskip -3.0truein
\caption[]{BR($\,(Z\rightarrow \mu^{\pm}\, e^{\pm})$) with respect
to the scale $1/R$ for $\rho=0.01$, $\bar{\xi}^{E}_{N,\tau mu}=1\,
GeV$, $\bar{\xi}^{E}_{N,\tau \mu}=1\,GeV$. Here the solid (dashed,
small dashed, dotted) line represents the BR without extra
dimension (with a single extra dimension, with two extra
dimensions where the leptons have non-zero Gaussian profiles in
the fifth extra dimension, with two extra dimensions where the
leptons have non-zero Gaussian profiles in both extra dimensions)}
\label{ZtaumuRr}
\end{figure}
\begin{figure}[htb]
\vskip -3.0truein \centering \epsfxsize=6.8in
\leavevmode\epsffile{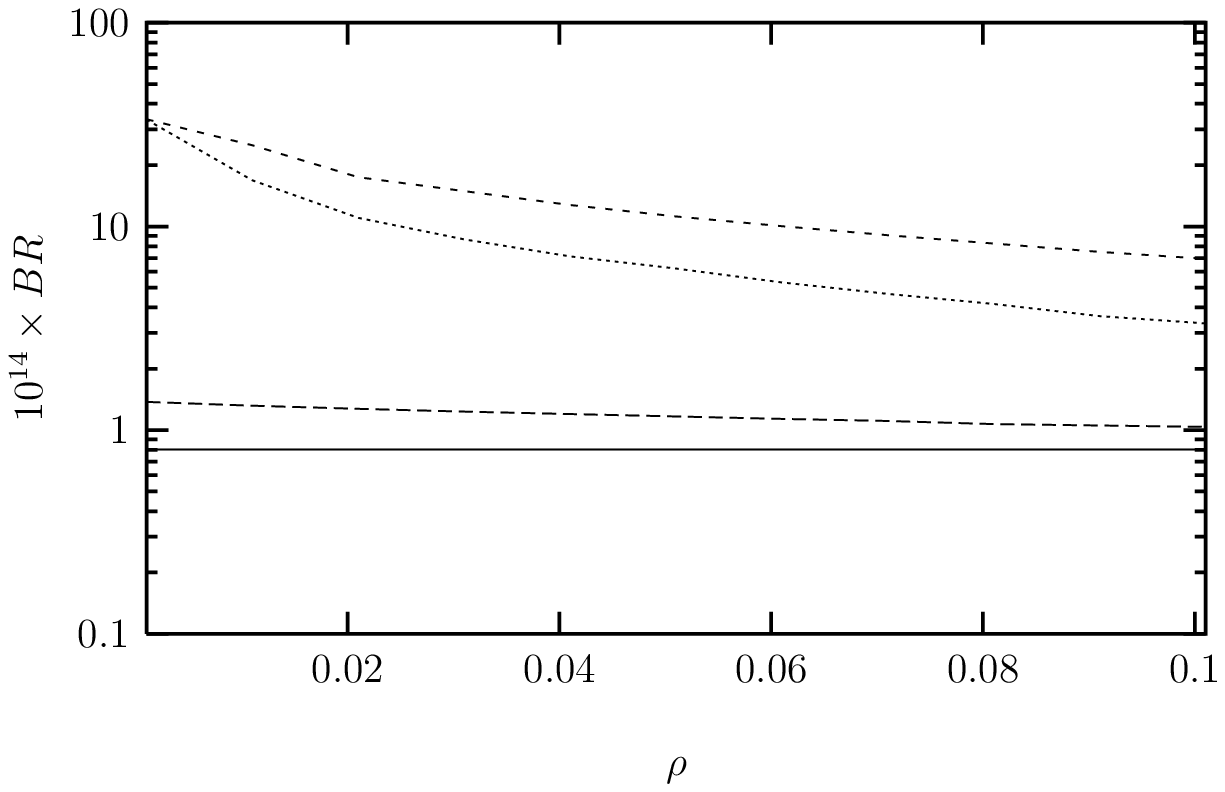} \vskip -3.0truein \caption[]{The
same as Fig. \ref{ZmueRr} but with respect to parameter $\rho$ and
for $1/R=500\, GeV$.} \label{Zmuero}
\end{figure}
\begin{figure}[htb]
\vskip -3.0truein \centering \epsfxsize=6.8in
\leavevmode\epsffile{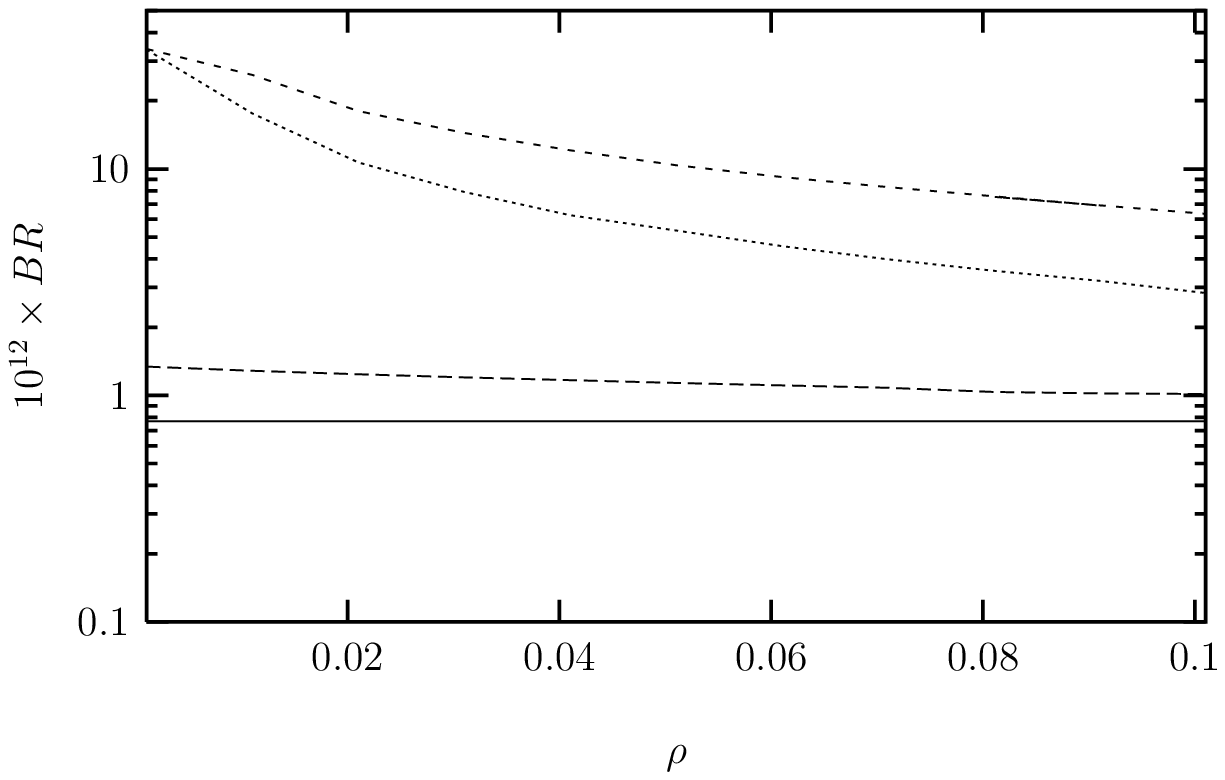} \vskip -3.0truein \caption[]{The
same as Fig. \ref{ZtaueRr} but with respect to parameter $\rho$
and for $1/R=500\, GeV$.} \label{Ztauero}
\end{figure}
\begin{figure}[htb]
\vskip -3.0truein \centering \epsfxsize=6.8in
\leavevmode\epsffile{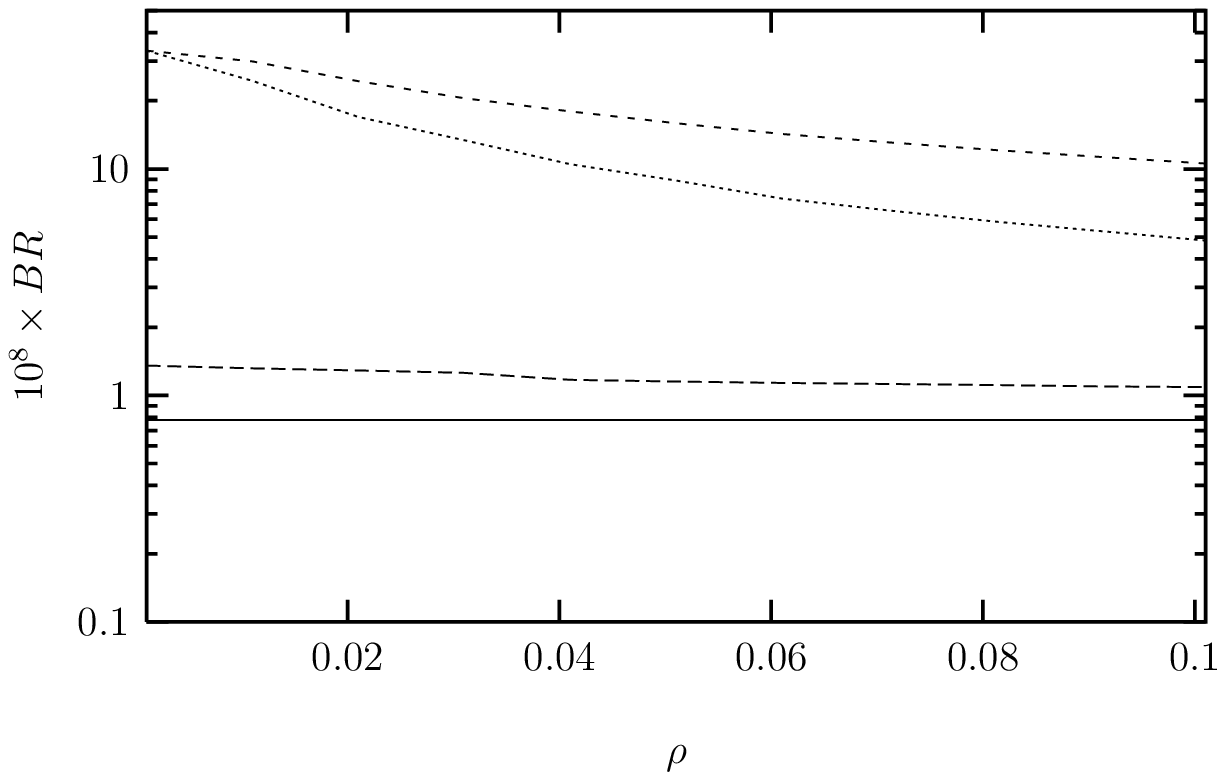} \vskip -3.0truein \caption[]{The
same as Fig. \ref{ZtaumuRr} but with respect to parameter $\rho$
and for $1/R=500\, GeV$. } \label{Ztaumuro}
\end{figure}
\end{document}